\documentstyle[aps,multicol]{revtex}
\begin{document}
\author{Zu-Jian Ying$^{\text{a)}}$, You-Quan Li$^{\text{a), b)}}$ 
and Shi-Jian Gu$^{\text{a)}}$}
\address{
$^{\text{a)}}$Zhejiang Institute of Modern Physics, Zhejiang
University, Hangzhou 310027, China\\
$^{\text{b)}}$Institut f\"ur Physik, Universit\"at Augsburg, D-86135
Augsburg, Germany}
\title{Symmetries in the Hubbard model with n-fold orbital degeneracy}
\date{Received: \qquad 2000}

\maketitle

\begin{abstract}
The present paper studies the symmeries of the Hubbard model of electrons
with generally $n$-fold orbital degeneracy. It's shown $SU_d(2n)$ and $%
SU_c(2n)$ symmetries hold respectively for the model with completely
repulsive or attractive on-site interaction and that with partly attractive
interactions. An extended Lieb-Mattis transformation is given to map these
two symmetries into each other. The sub-symmetry $SU_d^{(e)}(n)\otimes
SU_d^{(o)}(n)$ is found to be shared by the two models with arbitrary
chemical potential $\mu $. By assuming at most two electrons on each site
it's found that $SU_d(2n)$ and $SU_c(2n)$ both exist in each kind of the two
models and consequently lead to a larger symmetry $SU_d(2n)\times SU_c(2n)$.
Another underlying symmetry $\bigl( SU_c^{(e)}(2)_{{\cal P}}\times ...\times
SU_c^{(e)}(2)_{{\cal P}}\bigr) 
\otimes \bigl( SU_c^{(o)}(2)_{{\cal P}}\times ...\times SU_c^{(o)}(2)_{{\cal %
P}}\bigr)$ is also revealed for the unified $U$ model under the excluding.
The symmetry is valid for the partially attractive model with chemical
potential $\mu =-U.$


\vspace{10mm}\noindent
\end{abstract}


\section{Introduction}

Recently considerable attention has been directed to the studies on
correlated electrons in the present of orbital degree of freedom which is
relevant to transitional-metal oxides\cite{Oxides,Word,Bao,Pen,Feiner}, $%
C_{60}$ materials\cite{C60} and artificial quantum dot arrays\cite{QDots}.
Apart from the numerical\cite{Yama} and perturbative\cite{Azaria} works
theories based on symmetries were presented for one-dimensional models of
these systems. An $SU(4)$ theory describing spin systems with orbital
degeneracy was proposed\cite{Li98,Li99} for a theoretical understanding of
the observed unusual properties. The ground-state phase diagrams for the
system with a symmetry breaking of $SU(4)\rightarrow SU(2)\times SU(2)$ were
discussed in ref.\cite{Azaria,Itoi}. For the 2-fold orbital degenerate
Hubbard model a recent paper\cite{Li} presented the $SU(4)$ theory and
showed the underlying $SU_d(4)$ symmetry of spin-orbital double and a charge 
$SU_c(4)$ symmetry with an extended Lieb-Mattis transformation mapping those
two $SU(4)$ generators into each other. On the basis of elementary
degenerate perturbative theory, it was also shown that the effective
Hamiltonian is equivalent to the $SO(6)$ and $SU(4)$ Heisenberg models
respectively at half-filling and quarter-filling with strong coupling. In
ref.\cite{Li2000}, the one-dimensional $SU(4)$ Hubbard model is extensively
studied on the basis of Bethe ansatz solution. As for the symmetry theory of
one-dimensional Hubbard model without orbital degeneracy, it has been
well-investigated. Yang introduced the pairing operators and so constructed
the symmetry $SU(2)\times SU(2)$\cite{Yang}. Based on the symmetry Fabian
H.L.E$\beta $ler {\it et. al.}\cite{Fabian} discussed the completeness of
the Bethe ansatz solutions, M.Pernici\cite{Pernici} showed the off-diagonal
long-range order, D.B.Uglov and V.E.Korepin presented the Yangian symmetry $%
Y(sl(2))\oplus Y(sl(2))$ \cite{Uglov}, and Fabian H.L.E$\beta $ler and
Holger Frahm considered the density correlations\cite{Fabian2}. And it has
been argued that the two-dimensional single-band Hubbard model has
approximate $SO(5)$ symmetry\cite{Meixner}. But the research works on the
Hubbard model with orbital degeneracy are still in accumulation. In the
present paper we study the symmetries of the Hubbard model of electrons with
generally $n$-fold orbital degeneracy. We show and clarify that the $%
SU_d(2n) $ and $SU_c(2n)$ symmetries hold respectively, unlike in the simple
Hubbard model for which both of the two symmetries are valid, for model with
unified on-site interaction and that with partly attractive interactions.
But the sub-symmetry $SU_d^{(e)}(n)\otimes SU_d^{(o)}(n)$ is found both
valid for the two models and for arbitrary chemical potential $\mu $. An
extended Lieb-Mattis transformation as in \cite{Li} is given to map these
two symmetries into each other. By assuming at most two electrons on each
site, on the basis of which the Bethe ansatz can be applied in
one-dimensional model, we find the $SU_d(2n)$ and $SU_c(2n)$ symmetries both
exist in each kind of the two models so we have a larger symmetry $%
SU_d(2n)\times SU_c(2n)$. Under the exclusion another underlying symmetry $%
\bigl( SU_c^{(e)}(2)_{{\cal P}}\times ...\times SU_c^{(e)}(2)_{{\cal P}}%
\bigr)
\otimes \bigl( SU_c^{(o)}(2)_{{\cal P}}\times ...\times SU_c^{(o)}(2)_{{\cal %
P}}\bigr)$, which is not included in the $SU_d(2n)\times SU_c(2n)$, is also
revealed for the unified $U$ model with $\mu =U$ so that the model totally
possesses the symmetry $SU_d(2n)\times SU_c(2n)\times \bigl[\bigl( %
SU_c^{(e)}(2)_{{\cal P}}\times ...\times SU_c^{(e)}(2)_{{\cal P}}\bigr) %
\otimes \bigl( SU_c^{(o)}(2)_{{\cal P}}\times ...\times SU_c^{(o)}(2)_{{\cal %
P}}\bigr)\bigr]$. The underlying symmetry is also valid for the partially
attractive model with $\mu =-U.$

Consider the $n$-fold orbital degenerate electrons with states 
\begin{eqnarray}
\left| 1\right\rangle &=&\left| 1,\uparrow \right\rangle ,\quad \left|
2\right\rangle =\left| 1,\downarrow \right\rangle ,  \nonumber \\
&&\ ...,  \nonumber \\
\left| 2n-1\right\rangle &=&\left| n,\uparrow \right\rangle ,\quad \left|
2n\right\rangle =\left| n,\downarrow \right\rangle ,
\end{eqnarray}
where in a state $\left| l,\sigma \right\rangle $ $l$ denotes the $l$th
orbital component and $\sigma =\uparrow ,\downarrow $ label the spin
components. We start with a general Hamiltonian with n-fold orbital
degeneracy expressed by 
\begin{equation}
H=-\sum_{x,x^{\prime }}\sum_a\Bigl(t_{xx^{\prime }}C_a^{+}(x)C_a(x^{\prime
})+t_{xx^{\prime }}^{*}C_a^{+}(x^{\prime })C_a(x)\Bigr)+\sum_x\sum_{a\neq
a^{\prime }}U_{aa^{\prime }}n_a(x)n_{a^{\prime }}(x)-\mu \sum_{x,a}n_a(x)
\label{H}
\end{equation}
where $C_a^{+}(x)$ creates a fermion of state $\left| a\right\rangle $ at
site $x$ and $n_a(x)$ is the corresponding particle number operator. The
notation of site is not restricted to one-dimensional case.

\section{$SU_d(2n)$ and $SU_c(2n)$ symmetries}

Besides the $U(1)$ symmetry there exist two kinds of $SU(2n)$ symmetries for
the orbital degenerate Hubbard model. We define 
\begin{eqnarray}
E_{ss^{\prime }}&=&\sum_xC_s^{+}(x) C_{s^{\prime }}(x),  \nonumber \\
D_m&=&N_m-N_{m+1},\quad N_m=\sum_x C_m^{+}(x) C_m(x),
\end{eqnarray}
they fulfill the commutation relations 
\begin{eqnarray}
\lbrack E_{ss^{\prime }},E_{tt^{\prime }}]&=&\delta
_{s^{\prime},t}E_{st^{\prime}} -\delta _{s,t^{\prime }}E_{ts^{\prime}}, 
\nonumber \\
\lbrack D_m,E_{ss^{\prime}}]&=&( \delta _{m,s}-\delta _{m,s^{\prime}}
-\delta _{m+1,s}+\delta _{m+1,s^{\prime }}) E_{ss^{\prime}}.
\end{eqnarray}
These operators can construct an $SU(2n)$ Lie algebra 
\begin{equation}
SU_d(2n):\left\{ D_m,\quad E_{ss^{\prime }}\mid m=1,...,2n-1;1\leq s\neq
s^{\prime }\leq 2n\right\} ,  \label{SUd}
\end{equation}
with $n( 2n-1) $ of $E_{ss^{\prime }}$ and there are totally $(2n)^2-1$ of
generators. $\{D_m\}$ forms the commuting Cardan subalgebra of rank $2n-1$.
The $E_{s(s+1)}$'s are the generators related to the simple roots. For any
of $t_{xx^{\prime }}$ and $\mu $, all generators of $SU_d(2n)$ commute with
the Hamiltonian (\ref{H}) with unified on-site interaction $U_{aa^{\prime
}}=U$, so we have $SU_d(2n)$ symmetry of spin-orbital double in this case.

Let us define another set of operators 
\begin{eqnarray}
F_{\alpha _{2k-1}}=\sum_xf(x) C_{2k-1}^{+}(x) C_{2k}^{+}(x) ,  \nonumber \\
F_{\alpha _{2k}}=\sum_xf(x) C_{2k}(x) C_{2k+1}(x) ,  \nonumber \\
Q_m=\frac 12\sum_x [ C_m^{+}(x) C_m(x) +C_{m+1}^{+}(x) C_{m+1}(x) -1],
\end{eqnarray}
where $k=1,...,n$, $m=1,...,2n-1$, $f(x)^2=1$, and $f(x+\delta)=-f(x)$ for
any site $x$ and nearest-neighbor $x+\delta$. The above operators can
realize another $SU(2n)$ Lie algebra which we shall denote by $SU_c(2n)$: 
\begin{equation}
\left\{ Q_m,F_\alpha ,F_{-\alpha }\right\} .
\end{equation}
$\{Q_m\}$ is the Cardan subalgebra. The $F_{\alpha _{2k-1}}$and $F_{\alpha
_{2k}}$ are the generators related to the simple roots, other generators
relating to positive roots can be obtained by $F_{\alpha _i+\alpha
_j}=[F_{\alpha _i},F_{\alpha _j}]$, the generators with negative roots will
be $F_{-\alpha }=(F_\alpha )^{\dagger }$. If we assume that the on-site
coupling $U_{aa^{\prime}}=U$ for the states labeled by $a$, $a^{\prime}$
with different spin components while $U_{aa^{\prime }}=-U$ for states with
the same spin components 
\begin{eqnarray}
U_{aa^{\prime }} &=&U\text{ for odd-even pair }a,a^{\prime },  \nonumber \\
U_{aa^{\prime }} &=&-U\text{ for odd-odd or even-even pair }a,a^{\prime },
\label{Uaa'}
\end{eqnarray}
and the amplitudes $t_{xx^{\prime }}$ of odd-neighbor hopping are real and
those of even-neighbor hopping are imaginary 
\begin{eqnarray}
t_{xx^{\prime}}^{*} &=&t_{xx^{\prime}} \text{ when } x{\text-}x^{\prime}%
\text{ is odd neighbor},  \nonumber \\
t_{xx^{\prime}}^{*} &=&-t_{xx^{\prime}} \text{ when } x{\text -}x^{\prime}%
\text{ is even neighbor},  \label{OE}
\end{eqnarray}
we will have the following relations 
\begin{equation}
\lbrack H,F_{\alpha _{2k}}]=2(\mu-U)F_{\alpha _{2k}},\quad [H,F_{\alpha
_{2k-1}}]=-2(\mu-U)F_{\alpha _{2k-1}}.  \label{HSUc}
\end{equation}
If the chemical potential $\mu =U$, the Hamiltonian will commute with all
the generators of $SU_c(2n) $ so the model has the charge $SU_c(2n) $
symmetry. In terms of the partially attractive (\ref{Uaa'}), it can be
easily proved that such a Hamiltonian with $\mu =U$ has the half-filled form 
\begin{equation}
H=-\sum_{x,x^{\prime}}\sum_a\Bigl( t_{xx^{\prime}}C_a^{+}(x)C_a(x^{\prime})
+t_{xx^{\prime}}^{*}C_a^{+}(x^{\prime})C_a(x) \Bigr) +\sum_x\sum_{a\neq
a^{\prime}}U_{aa^{\prime}}\Bigl( n_a(x) -\frac{1}{2}\Bigr)
\Bigl(n_{a^{\prime}}(x)-\frac{1}{2}\Bigr).  \label{half}
\end{equation}
The usual Hubbard model with the nearest-neighbor hopping is included in the
class of (\ref{OE}). Whether $f(x)$ can be well defined depends on the
lattice structures, if in a lattice any sum of the nearest-neighbor pair $%
\delta +\delta^{\prime}$ is not a nearest-neighbor and there are even sites
on every perpendicular directions, $f(x)$ can be well defined as 
\begin{equation}
f(x) =\exp (i{\bf \pi }\cdot {\bf x}),\quad {\bf \pi} =( \pi ,\pi
,...),\quad {\bf x}=(x_1,x_2,...),
\end{equation}
$x_i$ is the $i$th components of the site coordinate in the lattice basis.
Such lattices (bipartite lattice) can be simple squared, centered squared in
2-dimension; simple cubic and body-centered cubic in 3-dimension. For 1-d
case and even total sites, $f(x) =\exp (i\pi x)$.

It should be noted that the $SU_d(2n)$ does not commute with the partially
attractive Hamiltonian which has only the $SU_c(2n)$ symmetry, neither does $%
SU_c(2n)$ with the unified $U$ model which possesses the $SU_d(2n)$
symmetry. The two kinds of symmetries can be mapped into each other by an
extended Lieb-Mattis transformation 
\begin{eqnarray}
C_a(x) &\mapsto &\exp (i{\bf \pi }\cdot {\bf x)}C_a^{+}( x) \text{ for even }%
a,  \nonumber \\
C_a(x) &\mapsto &C_a(x) \text{ for odd }a.  \label{LMT}
\end{eqnarray}
The transformation leaves the hopping term (\ref{OE}) invariant and changes
the sign of $U_{aa^{\prime}}$ with odd-even pair $a$,$a^{\prime}$ in (\ref
{half}) so maps the model (\ref{half}) into the unified $U_{aa^{\prime}}=-U$
model.

The particle number of each states and the total spin can be expressed by 
\begin{eqnarray}
N_{2n} &=& \Bigl( N_e-\sum_{m=1}^{2n-1}mD_m\Bigr)/2n,  \nonumber \\
N_j &=& \sum_{m=j}^{2n-1}D_m + \Bigl( N_e-\sum_{m=1}^{2n-1}mD_m\Bigr)/2n,
\quad j<2n,
\end{eqnarray}

\begin{equation}
S_{total}=\sum_{m=1}^{n}D_{2m-1} /2,
\end{equation}
where $N_e$ is the number of total electrons.

\section{$SU_d^{(e)}(n)\otimes SU_d^{(o)}(n)$ symmetry}

Unlike in the transitional Hubbard model for which both of the two
symmetries are valid for on-site attractive and repulsive interactions, as
we can see in previous section, for the orbital-degenerate Hubbard model the 
$SU_d(2n)$ symmetry of spin-orbital double only holds for the unified $U$
Hubbard model with arbitrary chemical potential $\mu $ whereas the charge $%
SU_c(2n)$ symmetry merely exists in the other partly-attractive half-filled
model of which the chemical potential is $\mu =U$. But considering that $%
SU_d(2n)$ and $SU_c(2n)$ share some common generators $E_{2\nu,2\nu^{%
\prime}} $ and $E_{2\nu -1,2\nu^{\prime}-1}$, we will find the shared
sub-symmetry $SU_d^{(e)}(n)\otimes SU_d^{(o)}(n)$ yields for the two models
with 
\begin{eqnarray}
SU_d^{(e)}(n):\; \left\{ D_{2\nu ,2\nu +2},
E_{2\nu^{\prime},2\nu^{\prime\prime}}|\nu \leq n-1,1\leq \nu^{\prime}\neq
\nu^{\prime\prime}\leq n\right\} ,  \nonumber \\
SU_d^{(o)}(n):\;
\left\{D_{2\nu-1,2\nu+1,E_{2\nu^{\prime}-1,2\nu^{\prime\prime}-1},} |\nu
\leq n-1,1\leq\nu^{\prime}\neq\nu^{\prime\prime}\leq n\right\} ,
\label{SUEO}
\end{eqnarray}
where $D_{m,m+2}=N_m-N_{m+2}$. Especially, we shall illustrate that the
symmetry is valid for the partially attractive model with arbitrary chemical
potential $\mu$.

As the hopping term in the Hamiltonian and the chemical potential term $%
\mu\sum_{x,a}n_a(x) $ are $SU_d(2n)$ invariant, whether the Hubbard model
possesses $SU_d(2n)$ symmetry depends on the commutation relation of the
on-site interacting term and the $SU_d(2n)$ generators 
\begin{eqnarray}
\Bigl[ E_m^k,\sum_{a\neq a^{\prime}}U_{aa^{\prime}}n_a(x) n_{a^{\prime}}(x)
\Bigr] &=& C_m^{+}(x)C_{m+k}(x) \Bigl(\sum_{a^{\prime}\neq
m+k}n_{a^{\prime}}(x)U_{m+k,a^{\prime}} -\sum_{a^{\prime}\neq
m}n_{a^{\prime}}(x)U_{m,a^{\prime}}\Bigr)  \nonumber \\
\,&\,& \;+\Bigl(\sum_{a\neq m+k}n_a(x) U_{a,m+k}-\sum_{a\neq m}n_a(x)
U_{a,m} \Bigr)C_m^{+}(x) C_{m+k}(x)  \label{commu}
\end{eqnarray}
where 
\[
E_m^k=\sum_xC_m^{+}(x) C_{m+k}(x). 
\]
Surely for the case of unified $U_{aa^{\prime}}=U$ we easily find that the
above commutation vanishes so that we have the $SU_d(2n)$ symmetry in this
case, as is obtained in the second section. Although for the partially
attractive model the above commutator does not go null for all $k^{\prime}$s
and consequently we do not have $SU_d(2n)$ symmetry, the case with even $k$
will be an exception. From the partially attractive (\ref{Uaa'}) we find for
even $k$'s 
\[
U_{a,m+k}=U_{a,m}=-U, 
\]
\begin{eqnarray*}
\Bigl(\sum_{a\neq m+k}n_a(x) U_{a,m+k}-\sum_{a\neq m}n_a(x) U_{a,m}\Bigr)
&=&n_m(x) U_{m,m+k}-n_{m+k}(x) U_{m+k,m} \\
&=&\bigl(n_m(x)-n_{m+k}(x)\bigr)(-U).
\end{eqnarray*}
Then eq.(\ref{commu}) becomes 
\begin{eqnarray*}
&\,& \Bigl[ E_m^k,\sum_{a\neq
a^{\prime}}U_{aa^{\prime}}n_a(x)n_{a^{\prime}}(x)\Bigr] \\
&=&C_m^{+}(x) C_{m+k}(x)\bigl( n_m(x) -n_{m+k}(x)\bigr)(-U) +\bigl(%
n_m(x)-n_{m+k}(x)\bigr)C_m^{+}(x) C_{m+k}(x)(-U) \\
&=&\bigl(-C_m^{+}(x) C_{m+k}(x) +C_m^{+}(x) C_{m+k}(x)\bigr)(-U)=0
\end{eqnarray*}
where we have used relations: $n_mC_m=C_m^{+}n_m=0$, $n_mC_m^{+}=C_m^{+}$, $%
C_mn_m=C_m$. $E_m^{k}$ with even $k$'s and odd (or even) $m$'s correspond to
the $SU_d^{(o)}(n)$ (or $SU_d^{(e)}(n)$) generators. As a result, the $%
SU_d^{(e)}(n)\otimes SU_d^{(o)}(n)$ generators in (\ref{SUEO}) commute with
the Hamiltonian of the partially attractive model for any chemical potential 
$\mu $. Therefore the $SU_d^{(e)}(n)\otimes SU_d^{(o)}(n)$ symmetry is
shared by the two models both for arbitrary $\mu$.

\section{At most two electrons on each site}

The application of Bethe ansatz method to the 1-dimensional degenerate
Hubbard model is based on such an assumption that prevents scattering
process involving three or more electrons on one site\cite
{Li2000,Sutherland,P}. For the traditional Hubbard model the configurations
of more than two electrons on one site are excluded automatically by the
Pauli principle. In the continuum limit and for small densities or $U\gg t$
in the lattice model, the unwanted configurations in degenerate Hubbard
model become negligible, so the Hamiltonian with three-electron
configurations excluded will describe the system well. If we exclude more
than two electrons on each site, we will find that the Hamiltonian has both $%
SU_d(2n)$ and $SU_c(2n)$ symmetries and furthermore a larger symmetry $%
SU_d(2n)\times SU_c(2n)$. In addition, we will find an underlying symmetry $%
\bigl( SU_c^{(e)}(2)_{{\cal P}}\times ...\times SU_c^{(e)}(2)_{{\cal P}}%
\bigr) \otimes \bigl( SU_c^{(o)}(2)_{{\cal P}}\times ...\times
SU_c^{(o)}(2)_{{\cal P}}\bigr) $.

Consider $U_{aa^{\prime }}=U$ case, the Hamiltonian reads 
\begin{equation}
{\cal H}=-t\sum_{<x,x^{\prime }>}\sum_a{\cal P}C_a^{+}(x)C_a(x^{\prime })%
{\cal P}+U\sum_x\sum_{a\neq a^{\prime }}n_a(x)n_{a^{\prime }}(x)-\mu
\sum_{x,a}n_a(x),  \label{PH}
\end{equation}
where $<x,x^{\prime }>$ represents the nearest neighbor sites and let us
define 
\begin{equation}
{\cal F}_{\alpha _{2k-1}}=\sum_x\exp (i{\bf \pi }\cdot {\bf x}){\cal P}%
C_{2k-1}^{+}(x)C_{2k}^{+}(x){\cal P},
\end{equation}
\begin{equation}
{\cal F}_{\alpha _{2k}}=\sum_x\exp (i{\bf \pi }\cdot {\bf x}){\cal P}%
C_{2k}(x)C_{2k+1}(x){\cal P},
\end{equation}
where the operator ${\cal P}$ projects onto the subspace of states having at
most two electrons on each site\cite{P}. Other generators can be given from
the ${\cal F}_{\alpha _{2k-1}}$ and ${\cal F}_{\alpha _{2k}}$ as in the $%
SU_c(2n)$ in section II. The ${\cal P}$ operator excludes such terms as $%
n_sC_{s^{\prime }}^{+}C_{s^{\prime \prime }}^{+}$, $C_sC_{s^{\prime
}}n_{s^{\prime \prime }}$ with three different $s,s^{\prime }$ and $%
s^{\prime \prime }$ so that 
\begin{equation}
\lbrack {\cal P}C_s^{+}(x)C_{s^{\prime }}^{+}(x){\cal P},\sum_{x^{\prime
}}\sum_{a\neq a^{\prime }}n_a(x^{\prime })n_{a^{\prime }}(x^{\prime })]=-2%
{\cal P}C_s^{+}(x)C_{s^{\prime }}^{+}(x){\cal P},
\end{equation}
thus we have the relations similar to (\ref{HSUc}) 
\begin{equation}
\lbrack {\cal H},{\cal F}_{\alpha _{2k}}]=2(\mu -U){\cal F}_{\alpha
_{2k}},\quad [{\cal H},{\cal F}_{\alpha _{2k-1}}]=-2(\mu -U){\cal F}_{\alpha
_{2k-1}}.
\end{equation}
Set $\mu =U$ and the ${\cal F}_{\alpha _{2k-1}}$ and ${\cal F}_{\alpha
_{2k}} $ will commute with the ${\cal H}$. ${\cal F}_{\alpha _{2k-1}}$, $%
{\cal F}_{\alpha _{2k}}$ and $N_s-N_{s^{\prime }}$ can generate the $%
SU_c(2n) $ symmetry so that ${\cal H}$ has both $SU_d(2n)$ and $SU_c(2n)$
symmetries, 
\begin{equation}
SU_c(2n):\left\{ {\cal Q}_m,\quad {\cal F}_\alpha \right\} ,  \label{PSUc}
\end{equation}
\begin{equation}
SU_d(2n):\left\{ {\cal D}_m,\quad {\cal E}_{ss^{\prime }}\right\} ,
\end{equation}
where ${\cal Q}_m={\cal P}Q_m{\cal P}$ and the $SU_d(2n)$ generators are
correspondingly revised to be 
\begin{equation}
{\cal E}_{ss^{\prime }}={\cal P}E_{ss^{\prime }}{\cal P},\quad {\cal D}_m=%
{\cal P}D_m{\cal P}.
\end{equation}
It also can be similarly shown that both of the two symmetries hold for the $%
{\cal P}$-modified model with partially attractive $U_{aa^{\prime }}=U$,$-U$%
. As the two symmetries both hold for each of the models we can construct
the larger symmetry $SU_d(2n)\times SU_c(2n)$ for each of them. If we do not
exclude more than two electrons on each site, these two symmetries
respectively belong to different models.

Besides $SU_d(2n)\times SU_c(2n)$ there exist some less obvious symmetries.
Define 
\begin{equation}
{\cal F}_{2k,2k^{\prime }}^{(+)}=\sum_x\exp (i{\bf \pi }\cdot {\bf x}){\cal P%
}C_{2k}^{+}(x)C_{2k^{\prime }}^{+}(x){\cal P},\quad {\cal F}_{2k,2k^{\prime
}}^{(-)}=({\cal F}_{2k,2k^{\prime }}^{(+)})^{\dagger },
\end{equation}
\begin{equation}
{\cal F}_{2k-1,2k^{\prime }-1}^{(+)}=\sum_x\exp (i{\bf \pi }\cdot {\bf x})%
{\cal P}C_{2k-1}^{+}(x)C_{2k^{\prime }-1}^{+}(x){\cal P},\quad {\cal F}%
_{2k-1,2k^{\prime }-1}^{(-)}=({\cal F}_{2k-1,2k^{\prime }-1}^{(+)})^{\dagger
},
\end{equation}
such generators are not included in (\ref{PSUc}) which contains pairing
operators ${\cal F}_{ss^{\prime }}^{(\pm )}$ only with odd-even pair $%
ss^{\prime }$. It can be easily verified that these operators also commute
with the unified $U$ model (\ref{PH}) with $\mu =U$ so we find an underlying
symmetry $(SU_c^{(e)}(2)_{{\cal P}}\times ...\times SU_c^{(e)}(2)_{{\cal P}%
})\otimes (SU_c^{(o)}(2)_{{\cal P}}\times ...\times SU_c^{(o)}(2)_{{\cal P}%
}) $ with 
\begin{equation}
SU_c^{(e)}(2)_{{\cal P}}:\left\{ {\cal Q}_{2k,2k^{\prime }},\quad {\cal F}%
_{2k,2k^{\prime }}^{(+)},\quad {\cal F}_{2k,2k^{\prime }}^{(-)}\right\}
\quad k\neq k^{\prime },\quad k,k^{\prime }=1,...,n,
\end{equation}
\begin{equation}
SU_c^{(o)}(2)_{{\cal P}}:\left\{ {\cal Q}_{2k-1,2k^{\prime }-1},\quad {\cal F%
}_{2k-1,2k^{\prime }-1}^{(+)},\quad {\cal F}_{2k-1,2k^{\prime
}-1}^{(-)}\right\} \quad k\neq k^{\prime },\quad k,k^{\prime }=1,...,n.
\end{equation}
There are respectively $C_n^2=n(n-1)/2$ of the $SU_c^{(e)}(2)_{{\cal P}}$
and $SU_c^{(o)}(2)_{{\cal P}}$ symmetries. The extended Lieb-Mattis
transformation (\ref{LMT}) maps the above symmetry into itself. An revised
Lieb-Mattis transformation mapping into the corresponding $SU_d^{(e)}(2)_{%
{\cal P}}$ and $SU_d^{(o)}(2)_{{\cal P}}$ will involve a third kind of
Hamiltonians with partially attractive $U_{aa^{\prime }}$, which differs
from what we discussed before. So finally we have the symmetry $%
SU_d(2n)\times SU_c(2n)\times \left[ (SU_c^{(e)}(2)_{{\cal P}}\times
...\times SU_c^{(e)}(2)_{{\cal P}})\otimes (SU_c^{(o)}(2)_{{\cal P}}\times
...\times SU_c^{(o)}(2)_{{\cal P}})\right] $ for (\ref{PH}) with $\mu =U$.
But for $U_{aa^{\prime }}=U$, $-U$ case the underlying symmetry is valid for
another chemical potential. The Hamiltonian under that exclusion is 
\begin{equation}
{\cal H}^{\prime }=-\sum_{<x,x^{\prime }>}\sum_at{\cal P}C_a^{+}(x)C_a(x^{%
\prime }){\cal P}+\sum_x\sum_{a\neq a^{\prime }}U_{aa^{\prime
}}n_a(x)n_{a^{\prime }}(x)-\mu \sum_{x,a}n_a(x)  \label{PH'}
\end{equation}
where the partially attractive interaction $U_{aa^{\prime }}$ is also
defined by (\ref{Uaa'}). Compared with (\ref{HSUc}) the commutation
relations are different 
\begin{equation}
\lbrack {\cal H}^{\prime },{\cal F}_{2k,2k^{\prime }}^{(+)}]=-2(\mu +U){\cal %
F}_{2k,2k^{\prime }}^{(+)},
\end{equation}
\begin{equation}
\lbrack {\cal H}^{\prime },{\cal F}_{2k,2k^{\prime }}^{(-)}]=2(\mu +U){\cal F%
}_{2k,2k^{\prime }}^{(-)}.
\end{equation}
The different sign of $U$ comes from $U_{2k,2k^{\prime }}=-U$ while in eq.(%
\ref{HSUc}) it's $U_{2k,2k\pm 1}=U$. Therefore the symmetry $\bigl( %
SU_c^{(e)}(2)_{{\cal P}}\times ...\times SU_c^{(e)}(2)_{{\cal P}}\bigr)
\otimes \bigl( SU_c^{(o)}(2)_{{\cal P}}\times ...\times SU_c^{(o)}(2)_{{\cal %
P}}\bigr) $ holds for $\mu =-U$. It should be noted the symmetry is valid
under the exclusion, without the ${\cal P}$-exclusion its generators will
commute with neither of the two kinds of models. And unlike $%
SU_d^{(o)}(n)\otimes SU_d^{(e)}(n)\subset SU_d(2n)$ in section III, none of
the $SU_c^{(o)}(2)_{{\cal P}}$ or $SU_c^e(2)$ is any sub-symmetry of $%
SU_c(2n)$.

Considering the commutations ${\cal F}_{\alpha _{2k}}$ and ${\cal E}%
_{ss^{\prime }}$, we have 
\begin{equation}
\lbrack SU_d^{(ss^{\prime })}(2),SU_c^{(ss^{\prime })}(2)]=0
\end{equation}
\begin{equation}
\lbrack SU_d^{(ss^{\prime })}(2),SU_c^{(s^{\prime \prime }s^{\prime \prime
\prime })}(2)]=0,\text{for seperate pairs (}ss^{\prime }\text{) and (}%
s^{\prime \prime }s^{\prime \prime \prime })\text{,}
\end{equation}
but 
\[
\bigl[ SU_d^{(ss^{\prime })}(2),SU_c^{(s^{\prime }s^{\prime \prime })}(2)%
\bigr]\neq 0, 
\]
where $SU_d^{(ss^{\prime })}(2)$ and $SU_c^{(ss^{\prime })}(2)$ are
sub-symmetries involving only the states $s$ and $s^{\prime }$. The whole
symmetry cannot be written in a direct product $SU_d(2n)\otimes SU_c(2n)$
but $SU_d(2n)\times SU_c(2n)$. This is also a difference from the single
band Hubbard model of which the symmetry in our notation is a direct product
of the two $SU(2)$'s: i.e., $SO(4)\simeq SU_d(2)\otimes SU_c(2)$. Therefore
for the one-dimensional model (\ref{PH}) which can be solved by the Bethe
ansatz\cite{P}, the discussion on the completeness of the Bethe ansatz
solution and the off-diagonal long-range order will be expected quite
different since in the single band case it's based on the vanishing
commuation of the $SU_d(2)$ and $SU_c(2)$.

\section{Brief summary}

In summary, we studied the symmetries of the Hubbard model of $n$-fold
orbital degenerate electrons. We show and clarify that the $SU_d(2n)$ and $%
SU_c(2n)$ symmetries hold respectively for the model with unified on-site
interaction and that with partly attractive interactions. An extended
Lieb-Mattis transformation is given to map these two symmetries into each
other. But the sub-symmetry $SU_d^{(e)}(n)\otimes SU_d^{(o)}(n)$ is found to
be possessed by the two models and both for arbitrary chemical potential $%
\mu $. By excluding more than two electrons on the same sites we find the $%
SU_d(2n)$ and $SU_c(2n)$ symmetries both exist in each kind of the two
models, so we have an enlarged symmetry $SU_d(2n)\times SU_c(2n)$. Under
this exclusion, another underlying symmetry $( SU_c^{(e)}(2)_{{\cal P}%
}\times ...\times SU_c^{(e)}(2)_{{\cal P}}) \otimes ( SU_c^{(o)}(2)_{{\cal P}%
}\times ...\times SU_c^{(o)}(2)_{{\cal P}}) $ is also found for the unified $%
U$ model with chemical potential $\mu =U$, and consequently this model has
the symmetry $SU_d(2n)\times SU_c(2n)\times \left[ ( SU_c^{(e)}(2)_{{\cal P}%
}\times ...\times SU_c^{(e)}(2)_{{\cal P}}) \otimes ( SU_c^{(o)}(2)_{{\cal P}%
}\times ...\times SU_c^{(o)}(2)_{{\cal P}}) \right] $. The underlying
symmetry is valid for the partially attractive model with chemical potential 
$\mu =-U.$

\section{Acknowledgments}

The work is supported by NSFC-19975040 and EYFC98 of China Education
Ministry, YQL is also supported by AvH Stiftung.


\begin{references}
\bibitem{Oxides}  D.B. McWhan, T.M. Rice, and J.P. Remeika, Phys. Rev. Lett. 
{\bf 23}, 1384 (1969); D.B. McWhan, A. Menth, J.P. Remeika, W.F. Brinkman,
and T.M. Rice, Phys. Rev. B {\bf 7,} 1920 (1973); C. Castellani, C.R.
Natoli, and J. Ranninger, {\it ibid}. {\bf 18}, 4945 (1978); {\bf 18}, 4967
(1978); {\bf 18}, 5001 (1978).

\bibitem{Word}  R.E. Word, S.A. Werner, W.B. Yelon, J.M. Honig, and S.
Shivashankar, Phys. Rev. B {\bf 23}, 3533 (1981); K.I. Kugel, and D.I.
Khomskii, Usp. Fiz Nauk. {\bf 136}, 621 (1982) [Sov. Phys. Usp. {\bf 25},
231 (1982)].

\bibitem{Bao}  W. Bao, C. Broholm, S.A. Carter, T.F. Rosenbaum, G. Aeppli,
S.F. Trevino, P. Metcalf, J.M. Honig, and J. Spalek, Phys. Rev. Lett. {\bf 71%
}, 766 (1993).

\bibitem{Pen}  H.F. Pen, J. Brink, D.I. Komomskii, and G.A. Sawatzky, Phys.
Rev. Lett. {\bf 78}, 1323 (1996); B.J. Sternlieb, J.P. Hill, and U.C.
Wildgruber, {\it ibid}. {\bf 76}, 2196 (1996); Y. Moritomo, A. Asamitsu,
H.Kuwahara, and Y. Tokura, Nature (London) {\bf 380}, 141 (1996).

\bibitem{Feiner}  L.F. Feiner, A.M. Oles, and J. Zaanan, Phys. Rev. Lett. 
{\bf 78}, 2799 (1997); S. Ishihara, J. Inoue, and S. Maekawa, Phys. Rev. B 
{\bf 55}, 8280 (9197); S. Ishihara, M. Yamanaka, and N. Nagaosa, {\it ibid}. 
{\bf 56}, 686 (1997); R. Shiina, T. Nishitani, and H. Shiba, J. Phys. Soc.
Jpn. {\bf 66}, 3159 (1997); K. Yamaura, M. Takano, A. Hirano, and R. Kanno,
J. Solid Stat. Chem. {\bf 127}, 109 (1997); H. Kawano, R. Kajimoto, H.
Yoshizawa, Y. Tomioka, H. Kuwahara, and Y. Tokura Phys. Rev. Lett. {\bf 76},
2196 (1997); Y. Okimoto, T. Katsufuji, T. Ishikawa, T. Arima, and Y. Tokura,
Phys. Rev. B {\bf 55}, 4206 (1997).

\bibitem{C60}  D.P. Arovas and A. Auerbach, Phys. Rev. B {\bf 52}, 10114
(1995).

\bibitem{QDots}  A. Onufric and J.B. Marston, Phys. Rev. B {\bf 59}, 12573
(1999).

\bibitem{Yama}  Y. Yamashita, N. Shibata, and K. Ueda, Phys. Rev. B {\bf 58}%
, 9114 (1998).

\bibitem{Azaria}  P. Azaria, E. Boulat, and P. Lecheminant, Phys. Rev. B 61
(2000) 12112. Y.L. Lee and Y.W. Lee, Phys. Rev. B {\bf 61}, 6765 (2000).

\bibitem{Li98}  Y.Q. Li, M. Ma, D.N. Shi, and F.C. Zhang, Phys. Rev. Lett. 
{\bf 81}, 3527 (1998).

\bibitem{Li99}  Y.Q. Li, M. Ma, D.N. Shi, and F.C. Zhang, Phys. Rev. B {\bf %
60}, 12781 (1999).

\bibitem{Itoi}  C. Itoi, S. Qin, and I. Affleck, Phys. Rev. B {\bf 61}, 6747
(2000); Y. Yamashita, N. Shibata, and K. Ueda, J. Phys. Soc. Jpn. {\bf 69},
242 (2000).

\bibitem{Li}  You-Quan Li and Ulrich Eckern, cond-mat/9911192, to be
published in Phys. Rev. B.

\bibitem{Li2000}  Y.Q. Li, S.J. Gu, Z.J. Ying, and U. Eckern, Phys.Rev.B 
{\bf 62,} 4860 (2000).

\bibitem{Yang}  C.N. Yang, Phys. Rev. Lett. {\bf 63}, 2144 (1989).

\bibitem{Fabian}  Fabian H.L. E$\beta $ler, Vladimir E. Korepin and Kareljan
Schoutens, Nucl. Phys. B {\bf 384}, 431 (1992).

\bibitem{Pernici}  M. Pernici, Europhys. Lett. {\bf 12}, 75 (1990).

\bibitem{Uglov}  D.B. Uglov and V.E. Korepin, Phys. Lett. A {\bf 190}, 238
(1994).

\bibitem{Fabian2}  Fabian H.L. E$\beta $ler and Holger Frahm, Phys. Rev. B 
{\bf 60}, 8540 (1999).

\bibitem{Meixner}  S. Merxner, W. Hanke, E. Demler, and S.C. Zhang, Phys.
Rev. Lett. {\bf 79}, 4902 (1997).

\bibitem{Sutherland}  B. Sutherland, {\it Exactly Solvable Problems in
Condensed Matter and Relativistic Field Theory}, edited by B. S. Shastry,
S.S. Jha and V. Singh (Springer, Berlin, 1985) pp. 1-95.

\bibitem{P}  T.C. Choy and F.D.M. Haldane, Phys. Lett. A {\bf 90}, 83
(1982); K. Lee and P. Schlottmann Phys. Rev. Lett. {\bf 63}, 2299 (1989),
Physica B {\bf 63}, 398 (1990); P. Schlottmann, Phys. Rev. B {\bf 43}, 3101
(1991); H. Frahm and A. Schadschneider, J. Phys. A: Math. Gen. {\bf 26},
1463 (1993).
\end{references}
\end{document}